# Unfolding large-scale online collaborative human dynamics

YilongZha[1,2], Tao Zhou[1,3,4*], Changsong Zhou[2,3*]

[1]CompleX Lab, Web Sciences Center, University of Electronic Science and Technology of China, Chengdu 611731, People's Republic of China.

[2]Department of Physics, Centre for Nonlinear Studies and The Beijing-Hong Kong-Singapore Joint Centre for Nonlinear and Complex Systems (Hong Kong), Institute of Computational and Theoretical Studies, Hong Kong Baptist University, Kowloon Tong, Hong Kong.

[3]Beijing Computational Science Research Center, Beijing 100084, People's Republic of China.

[4]Big Data Research Center, University of Electronic Science and Technology of China, Chengdu 611731, People's Republic of China.

*Correspondence to: zhutou@ustc.edu or cszhou@hkbu.edu.hk

**Abstract**: Large-scale interacting human activities underlie all social and economic phenomena, but quantitative understanding of regular patterns and mechanism is very challenging and still rare. Self-organized online collaborative activities with precise record of event timing provide unprecedented opportunity. Our empirical analysis of the history of millions of updates in Wikipedia shows a universal double power-law distribution of time intervals between consecutive updates of an article. We then propose a generic model to unfold collaborative human activities into three modules: (i) individual behavior characterized by Poissonian initiation of an action, (ii) human interaction captured by a cascading response to others with a power-law waiting time, and (iii) population growth due to increasing number of interacting individuals. This unfolding allows us to obtain analytical formula that is fully supported by the universal patterns in empirical data. Our modeling approaches reveal "simplicity" beyond complex interacting human activities.

**One Sentence Summary:** Simple generic model predicts universal double power-law distribution of interevent time in collaborative editing of articles in Wikipedia.

Quantitative understanding of regular patterns in human dynamics is of great importance but fair challenging because they are driven by complex decision-making processes, involving competing choices under limited time and cost, interactions with and influences from social peers, environmental effects, and so on. Previously, when detailed and precise records of human activities were seldom, individual activities were assumed to follow random Poissonian processes with exponential distributions of inter-event times (*1*). In contrast, recent experiments on deliberate human behaviors, such as communication through emails (*2*), surface mails (*3*), cell phones (*4*), instant messages (*5*) and text messages (*6*), and online activities including web browsing (*7*), movie watching (*8*), searching (*9*) and shopping (*10*), showed that individual activities usually embody the bursty nature featured by fat-tailed, power-law-like distributions of inter-event times. Several models have been proposed to explain the possible mechanisms, including task competition model driven by individual decision (*2,11,12*), non-stationary Poisson process driven by daily and weekly circadian circles (*13,14*) and adaptive interest (*15*). Besides the complexity in individual decision-making, these works do not take the interaction between individuals into explicit consideration. The interaction was theoretically explored by placing task competition model into a network of two (*16*) or more agents (*17*). Only recently, it was explicitly demonstrated that interaction can lead to a bimodal distribution of the inter-event times in short message communication which is mainly a two-person interaction system (*18*). Thus far, how interactions happen in a real system organized by a large number of individuals is still unknown. Modern information technology has allowed many new types of human interacting activities, in particular, lots of online communities emerge in the Internet world to perform collaborative activities, such as distributed collaborative writing based on wikis (*19*) and globalized development of software (*20*), where interactions between hundreds or even thousands of participants are widely found and the event times are precisely recorded, providing unprecedented opportunity to uncover statistical regularities and to reveal underlying mechanisms in a quantitative way. Thus far, previous works on human collaborative systems have mainly focused on the network structural analysis and vandalism detection (*21-24*), while the temporal patterns are less considered.

Wikipedia is the world's largest wiki system and it precisely records all online collaborative activities. Different from the collaborative development of software and other actions under regulation and control, Wikipedia is a typical self-organized system without centralized management. We analyzed millions of updating records of hundreds of articles in Wikipedia and found that the inter-update time distribution follows a universal double power-law form. Namely, the 'head' and 'tail' of such distribution are both power-law-like, but with different exponents. Strikingly, we showed that this universal yet complicated distribution can be quantitatively predicted by an analytical formula via unfolding the seemingly complex collaborative patterns into three generic modules related to individual behavior, interaction among individuals and growth of interacting population, respectively.

The updating records in Wikipedia (see S1 in *Supplementary Materials*(SM)) show universal temporal patterns in articles' evolutionary process. In large percentage of articles, the number of updates in an article, *N*, grows in an exponential form in its early stage and then saturates (Fig. 1A). A profound double power-law distribution $P(\tau)$ is observed, where $\tau$ is the time interval between two consecutive updates of an article. This distribution has a power-law 'head' and a power-law 'tail', with different exponents (Fig. 1B). Note that, such pattern is insensitive to articles' contents, updating rates, total update numbers and lasting time (see S2 and Fig. S1 in SM for all 625 articles with>5000 updates). As we will show later, our model can also quantitatively explain a few exceptions with irregular growing patterns.

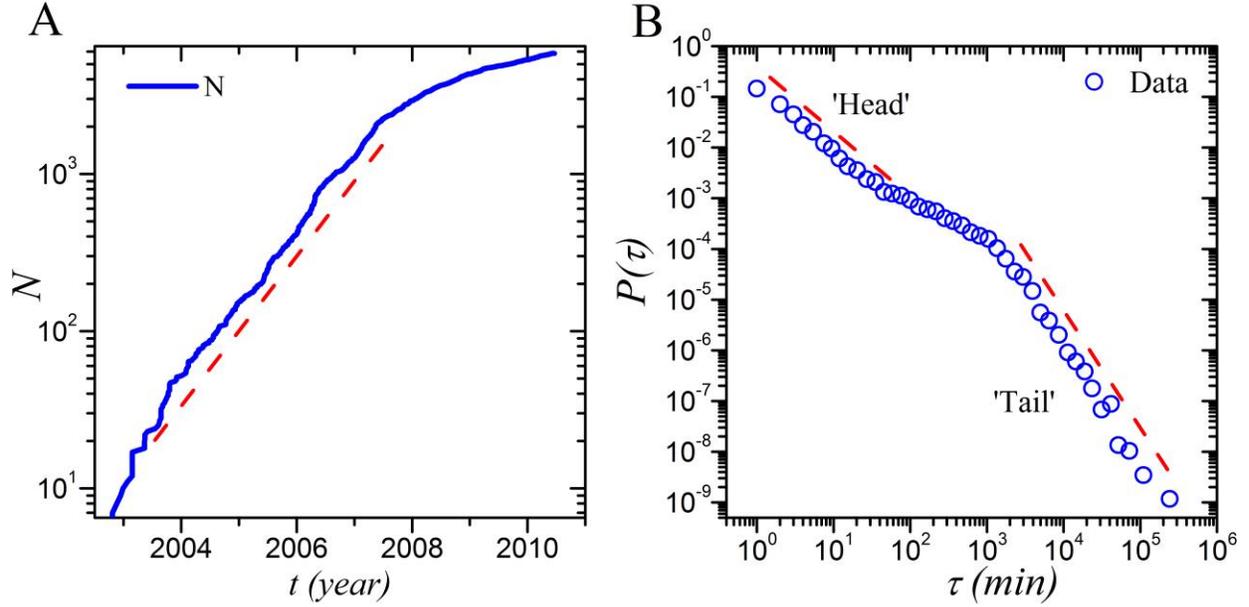

**Fig. 1**. **Regular patterns in the updating process of a typical Wikipedia article.** An example article 'Wikipedia' is used to illustrate the growing and distribution patterns. (**A**) Growth of the number of updates $N$ (blue). The eyes-guiding dashed line shows that the growth can be reasonably described by an exponential function. (**B**) Distribution of the inter-update time $\tau$, with the eyes-guiding dashed lines showing different slopes of the 'head' and 'tail'.

Collaborative activities typically happen in an evolving dynamical system consisting of many interacting individuals. To understand the major factors responsible for the observed regular patterns, we need to identify the dynamical features of individuals, the characteristics of interactions and the evolution of the whole system. In the collaborative editing of an article in Wikipedia, there are generally two types of updates: to initiate an update or to respond to others. Responsive updates are strongly related to the earlier ones (e.g., reverting the last update), typically with short waiting time from the last updates. Some updates are rather independent (e.g., adding a new section), typically with relatively longer time interval from the last updates and may further lead to a series of responsive updates. These two types of updates are respectively named as random initiations (I-type events) and responses (R-type events). Here, we propose to unfold a typical collaborative system into the following three generic modules.

(i) *Individual Behavior*. Initiations by individuals are assumed to be largely independent of other events, characterized by a random Poissonian process with an initiation rate $\lambda$. That is to say, at each time step, with probability $\lambda$, an I-type event happens. $\lambda$ increases with the number of participants.

(ii) *Interaction*. Responses can be induced by either type of earlier events. They are characterized by a cascading process with branching rate $a < 1$, namely each event has a chance $a$ to induce a responsive event. Inspired by the known empirical observations (*2,3,6,12,25-28*), the interval time between an event and its possibly induced event (i.e., waiting time) follows a power-law distribution

$$q(\tau) = \tau^{-\beta}/\zeta(\beta), \qquad (1)$$

where $\tau = 1,2,3,...$ denote discrete time intervals, and the normalizing factor $\zeta(\beta) = \sum_{i=1}^{\infty} i^{-\beta}$ is the Riemann zeta function.

(iii) *Population Growth*. Growth of population of participants leads to the growth of events through modules (i) and (ii). We found that the number of events $N$ (i.e., updates) is linear proportional to the number of participants (i.e., editors), thus we only consider the growing patterns of $N(t)$. Typically the growing rate $r(t) = dN(t)/dt$ is proportional to the size $N$, resulting in an exponential form $N(t) \sim e^{\alpha t}$, until some saturation sets in.

The analytical solution shows that the combination of modules (i) and (ii) with a constant initiation rate $\lambda$ will lead to a bimodal distribution of a power-law head and an exponential tail (Fig. 2A) (see S3 in SM)

$$P_\lambda(\tau) = e^{-\frac{\lambda}{1-a}\sum_{t=0}^{\tau-2} K(t)} K(\tau-1) - e^{-\frac{\lambda}{1-a}\sum_{t=0}^{\tau-1} K(t)} K(\tau), \qquad (2)$$

where $K(\tau) = 1 - a\zeta_\tau(\beta)/\zeta(\beta)$ and $\zeta_\tau(\beta) = \sum_{t=0}^{\tau} t^{-\beta}$. The 'head' at small $t$ has the same form $a\tau^{-\beta}/\zeta(\beta)$ as the waiting time in Eq.1, and the asymptotic 'tail' $(1-a) \cdot \lambda(1-\lambda)^{\tau-1}$ is an exponential curve resulted from the Poissonian process, mainly contributed by the I-type events. The exponential 'tail' shifts rightwards when $\lambda$ gets smaller (see Fig. 2A).

Since each initiation will induce a sequence of branching process with probability $a$, the expected number of updates induced by each I-type event is $S = 1/(1-a)$. The growth rate of the events is then $r = \lambda S$, leading to a theoretical relation between initiation and growth rates as:

$$\lambda = (1-a)r. \qquad (3)$$

Considering the population growth in module (iii), $\lambda$ becomes inhomogeneous and follows a distribution $\rho(\lambda)$, and the distribution of $\tau$ is

$$P(\tau) = \int \rho(\lambda) P_\lambda(\tau) d\lambda. \qquad (4)$$

For the exponential growth $N(t) \sim e^{\alpha t}$, it is obvious that $\rho(\lambda) = 1/(\lambda_2 - \lambda_1)$ is a uniform distribution between $\lambda_1 = (1-a)\alpha N_1$ and $\lambda_2 = (1-a)\alpha N_2$, which are determined (following Eq. 3, $r = \alpha N$), by the population size of the starting $N_1$ and ending $N_2$ of the exponential growth. Applying this uniform distribution to Eq. 4, it arrives at a very complicated analytical solution (see Eq. S6 in SM), which is actually a double power-law distribution with a power-law 'head' $a\tau^{-\beta}/\zeta(\beta)$ and a power-law 'tail' $(1-a)/(\lambda_2 - \lambda_1) t^{-2}$ in the range $1/\lambda_2 < \tau < 1/\lambda_1$, and the universal exponent -2 for the 'tail' is independent of the growing exponent $a$. In spite of the complicated form, to our knowledge, this is the first analytical solution based on a mechanistic model for the observed double power laws (see also some other complex systems exhibiting double power laws (*29,30*)). Fig. 2B compared the analytical solution with the simulation of an exponential growth process (see S4 in SM for simulation details), showing prefect consistence.

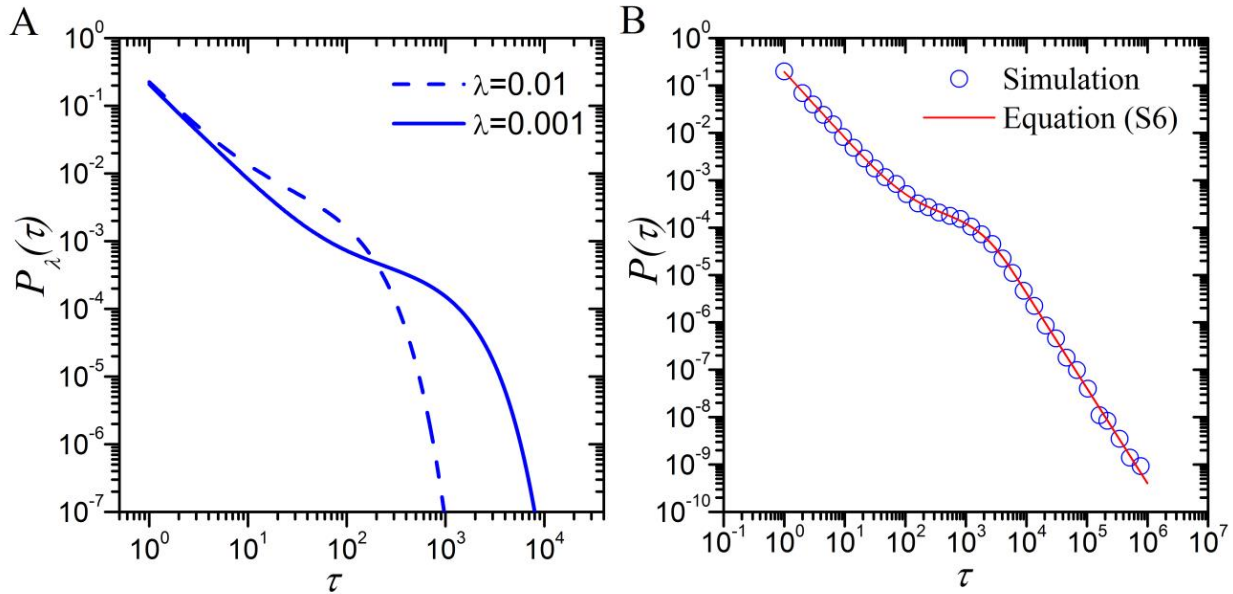

**Fig. 2. Bimodal and double power-law distributions from the model.** (**A**) The bimodal distribution at constant $\lambda$ (Eq.2). (**B**) The double power-law distribution (Eq. S6 in SM) with $\lambda_1 = 10^{-7}$ and $\lambda_2 = 10^{-3}$, compared to the result from a simulated growing process from $\lambda_1$ to $\lambda_2$. Other parameters are $b = 1.3$ and $a = 0.6$.

Our model assumptions and analytical solutions are fully supported by the empirical data from Wikipedia.

(i) *Universal power-law distribution of waiting time.* We have identified a large portion of R-type events from the history record of the 625 articles (see S5 in SM for how to distinguish I-type and R-type events). The waiting time for each article, as the model assumed, universally follows a power-law distribution, with the exponent $\beta$ narrowly distributed around $\overline{\beta} = 1.55$ (see a typical example in Fig. 3A and results for all the 625 articles in Fig. S3A and Fig. S4A).

(ii) *Bimodal distribution in the data segments with nearly constant growth rate.* We have designed an algorithm to find segments with nearly constant initiation rates (see S6 in SM). Altogether, 334 segments, each contains more than 1000 updates, were selected from 1948 articles. The inter-update time distribution for each segment appears to be bimodal (see Fig. 3B). We fit Eq. 2 to each segment by the maximum likelihood estimation, and 95.81% of the fittings can pass the Kolmogrov-Smirnov (KS) test with a significance level of 5% (see Fig. S7A and

Fig.S7B in SM for the statistics of the 334 segments). The parameters $\beta$ and $a$ distribute narrowly around their mean values $\bar{\beta} \approx 1.429$ and $\bar{a} \approx 0.605$, and could be regarded as independent of $r$ (see Fig. S6 and S6 in SM for the statistics and a possible explanation to the small discrepancy of $\beta$ in the above two very different experiments). Fig. 3C shows excellent consistence of the theory (Eq. 3) with empirical data. In a word, empirical analysis validates our model and demonstrates that the bimodal inter-event time distribution is a signature of a stable collaborative multi-individual system without variation of activity level (*18*).

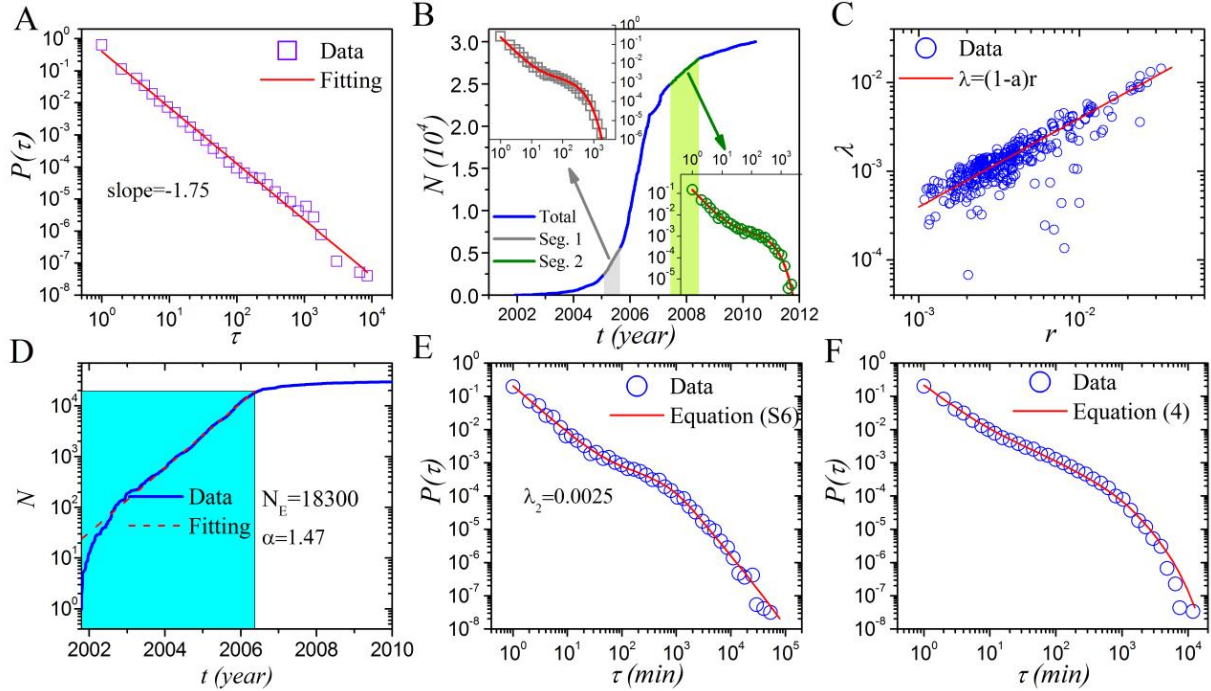

**Fig. 3. Verifying the model with data.** (**A**) Power-law distribution of waiting time extracted from responsive updates of the article 'Wikipedia'. (**B**) The growth of the number of updates $N(t)$ of the article 'Wikipedia'. Two linear segments (Seg. 1 and Seg. 2) were selected, each contains 3000 updates. Two insets show the bimodal distributions of $\tau$ for the two segments (symbols) and the corresponding theoretical predictions by Eq. 2 (red lines). (**C**) The estimated parameter $\lambda$ as a function of the growth rate $r$ from the 334 linear segments, compared to the theoretical prediction by Eq. 3 with $a = \bar{a} \approx 0.605$. (**D**) $N(t)$ of 'Wikipedia' (blue line) along with the fitting of an exponential function $\sim e^{\alpha t}$ (red dashed line) to the range $N < N_E$ (cyan area). (**E**) Theoretical double-power distribution (Eq. S6, red line) versus real data (blue circles) with parameters estimated from the cyan area in **D**. The parameters for theoretical formula are estimated using the maximum likelihood estimation. (**F**) The fitting of distribution of data from non-exponential part ($N > N_E$) by Eq. 4, where $\rho(\lambda)$ was estimated from the data.

(iii) *Double power-law distribution.* To compare the analytical solution with empirical data, we first identify the area $N < N_E$ with exponential growth (e.g., Fig. 3D) and the exponent

$\alpha$ (see S7 in SM for the method). The distribution $P(\tau)$ of the above exponentially growing part is fitted by the analytical solution Eq. S6 with parameters $a$, $\beta$ and $l_2$ estimated by the maximum likelihood method (see Fig. 3E), and 80.77% of the samples have passed K-S test of the double power-law hypothesis (see details in Fig. S7C, Fig. S7D and Fig. S8in SM). For the remaining part of the article ($N>N_E$), Eq.4 is fitted to the empirical distribution (see Fig. 3F), in which the distribution of initiation rate $\rho(\lambda)$ is obtained by numerical estimation of the growth rate $r$ from the data via Eq. 3.The theory is very robust: the fittings to most articles are as good as Fig. 3E and Fig. 3F, even including articles whose growing processes do not contain considerable exponential part (see Fig. S3 and Fig. S4 for the fittings to all the 625 articles, and S7C for the discussion on a universal fitting with $\bar{\beta}$ and $\bar{a}$ estimated from bimodal distribution).

Wikipedia sets up simple rules but displays complex phenomena that are largely originated from extensive human interactions. In spite of its high complexity, we show that such a system could still be unfolded into several simple and accessible components. Indeed, the temporal patterns of editing for disparate articles can be analytically reproduced with surprisingly high precision by the integration of three generic modules that respectively characterize the individual behavior, the interaction among individuals and the population growth. The proposed model is highly applicable: on one hand, it is essentially a parameter-free model since we do NOT need to adjust or optimize any parameters but every parameter is determined by the real data; on the other hand, it is a generic model for collaborative dynamics since all its mechanisms do NOT depend on the detailed rules or particular structures of Wikipedia. For example, e-mail communication is another typical collaborative system, according to the present theory, the corresponding initiation rate $\lambda_2$ is large, resulting in a transition from double power-law to single power-law in inter-event time distribution (see Fig. S9 in SM), which is also in accordance with the previous observations (*2,12*).

**References and Notes:**

1. F. A. Haight, *Handbook of the Poisson Distribution* (1967), pp. 100-107.

2. A. L. Barabási, The origin of bursts and heavy tails in human dynamics. *Nature* **435**, 207–211 (2005).

3. J. G. Oliveira, A. L. Barabási, Human dynamics: Darwin and Einstein correspondence patterns. *Nature* **437**, 1251-1251 (2005).

4. J. Candia, *et al*. Uncovering individual and collective human dynamics from mobile phone records. *J. Phys. A* **41**, 224015 (2008).

5. J. Leskovec, E. Horvitz, Planetary-scale views on a large instant-messaging network. *Proceeding of the 17th international conference on World Wide Web*, 915–924 (2008).

6. Z.-D. Zhao, H. Xia, M.-S. Shang, T. Zhou, Empirical Analysis on the Human Dynamics of a Large-Scale Short Message Communication System. *Chin. Phys. Lett.* **28**, 068901 (2011).

7. Z. Dezsö, *et al*. Dynamics of information access on the web. *Phys. Rev. E* **73**, 066132 (2006).

8. F. Radicchi, Human activity in the web. *Phys. Rev. E* **80**, 026118 (2009).

9. T. Zhou, H. A. T. Kiet, B. J. Kim, B. H. Wang, P. Holme, Role of activity in human dynamics. *Europhys. Lett.* **82**, 28002 (2008).

10. Z.-D. Zhao, *et al*. Emergence of scaling in human-interest dynamics. *Scientific Reports* **3**, 3472 (2013).

11. A. Vazquez, Exact results for the Barabási model of human dynamics. *Phys. Rev. Lett.* **95**, 248701 (2005).

12. A. Vazquez, *et al*. Modeling bursts and heavy tails in human dynamics. *Phys. Rev. E* **73**, 036127 (2006).

13. R. D. Malmgren, D. B. Stouffer, A. E. Motter, L. A. N. Amaral, A Poissonian explanation for heavy tails in e-mail communication. *Proc. Natl. Acad. Sci. U.S.A.* **105**, 18153–18158 (2008).

14. R. D. Malmgren, D. B. Stouffer, A. S. L. O. Campanharo, L. A. N. Amaral, On universality in human correspondence activity. *Science* **325**, 1696–1700 (2009).

15. X. Han, T. Zhou, B. Wang, Modeling human dynamics with adaptive interest. *New J. Phys.* **10**, 073010 (2008).

16. J. G. Oliveira, A. Vazquez, Impact of interactions on human dynamics. *Physica A* **338**, 187–192 (2009).

17. B. Min, K.I. Goh, I. M. Kim, Waiting time dynamics of priority-queue networks. *Phys. Rev. E* **79**, 056110 (2009).

18. Y. Wu, C. Zhou, J. Xiao, J. Kurths, H. J. Schellnhuber, Evidence for a bimodal distribution in human communication. *Proc. Natl. Acad. Sci. U.S.A.* **107**, 18803–18808 (2010).

19. C. Wei, B. Maust, J. Barrick, E. Cuddihy, J. H. Spyridakis, Wikis for supporting distributed collaborative writing. *Proceedings of the Society for Technical Communication 52nd Annual Conference* 204–209 (2005).

20. C. R. B. De Souza, D. Redmiles, P. Dourish, "Breaking the code", moving between private and public work in collaborative software development. *Proceedings of the 2003 International ACM SIGGROUP Conference on Supporting Group Work* 105–114 (2003).

21. A. Kittur, B. Suh, B. A. Pendleton, E. H. Chi, He says, she says: conflict and coordination in Wikipedia. *Proceedings of the SIGCHI Conference on Human Factors in Computing Systems* 453–462 (2007).

22. V. Zlatić, M. Božičević, H. Štefančić, M. Domazet, Wikipedias: Collaborative web-based encyclopedias as complex networks. *Phys. Rev. E* **74**, 016115 (2006).

23. V. Zlatić, H. Štefančić, Model of Wikipedia growth based on information exchange via reciprocal arcs. *Europhys. Lett.* **93**, 58005 (2011).

24. J. Török, et al. Opinions, Conflicts, and Consensus: Modeling Social Dynamics in a Collaborative Environment. *Phys. Rev. Lett.* **110**, 088701 (2013).

25. N. Li, N. Zhang, T. Zhou, Empirical analysis on temporal statistics of human correspondence patterns. *Physica A* **387**, 6391 (2008).

26. S. Wuchty, B. Uzzi, Human Communication Dynamics in Digital Footsteps: A Study of the Agreement between Self-Reported Ties and Email Networks. *PLoS ONE* **6**, e26972 (2011).

27. R. Hidalgo, A. Cesar, Conditions for the emergence of scaling in the inter-event time of uncorrelated and seasonal systems. *Physica A* **369**, 877–883 (2006).

28. Z.-Q. Jiang, W.-J. Xie, M.-X. Li, B. Podobnik, W.-X. Zhou, H. E. Stanley, Calling patterns in human communication dynamics. *Proc. Natl. Acad. Sci. U.S.A.* **110**, 1600–1605 (2013).

29. W. J. Reed, The Pareto, Zipf and other power laws. *Economics Letters* **74**, 15–19 (2001).

30. A. Corral, Local distributions and rate fluctuations in a unified scaling law for earthquakes. *Phys. Rev. E* **68**, 035102 (2003).

**Acknowledgments**: This work was partially supported by HKBU Strategic Development Fund and the Hong Kong Research Grant Council (HKBU12302914), and National Natural Science Foundation of China under Grant Nos. 11222543, 11275027 and 61433014.